\documentclass[preprint,review,10pt,sort&compress]{elsarticle}
\usepackage[margin=1 in]{geometry}
\usepackage{multirow}
\usepackage{xcolor}
\definecolor{UW}{RGB}{64, 38, 96}
\usepackage{amssymb}
\usepackage{float}

\journal{Composites Part B}

\begin{document}

\begin{titlepage}

\clearpage\thispagestyle{empty}

%\noindent {\footnotesize {{\em

%\hfill To be submitted to Cement and Concrete Composites} }} \\

\noindent

\hrulefill

\begin{figure}[h!]

\centering

\includegraphics[width=1.5 in]{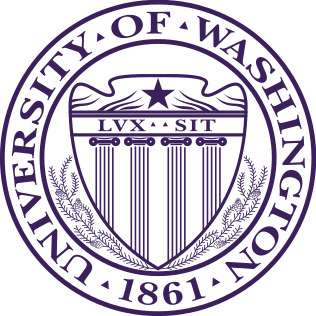}

\end{figure}

\begin{center}

{\color{UW}{

{\bf A\&A Program in Structures} \\ [0.1in]

William E. Boeing Department of Aeronautics and Astronautics \\ [0.1in]

University of Washington \\ [0.1in]

Seattle, Washington 98195, USA

}

}

\end{center} %\vskip 5mm

\hrulefill \\ \vskip 2mm

\vskip 0.5in

\begin{center}

{\large {\bf Micro-Computed Tomography Analysis of Damage in Notched Composite Laminates Under Multi-Axial Fatigue}}\\[0.5in]

{\large {\sc Yao Qiao, Marco Salviato}}\\[0.75in]

{\sf \bf INTERNAL REPORT No. 19-10/04E}\\[0.75in]

\end{center}

\noindent {\footnotesize {{\em Submitted to Composites Part B: Engineering \hfill October 2019} }}

\end{titlepage}

\newpage

\begin{frontmatter}


\cortext[cor1]{Corresponding Author, \ead{salviato@aa.washington.edu}}

\title{Micro-Computed Tomography Analysis of Damage in Notched Composite Laminates Under Multi-Axial Fatigue}


\author[address]{Yao Qiao}
\author[address]{Marco Salviato\corref{cor1}}

\address[address]{William E. Boeing Department of Aeronautics and Astronautics, Guggenheim Hall, University of Washington, Seattle, Washington 98195-2400, USA}

\begin{abstract}
\linespread{1}\selectfont

The broad application of polymer composites in engineering demands the deep understanding of the main damage mechanisms under realistic loading conditions and the development of proper physics-based models.

Towards this goal, this study presents a comprehensive characterization of the main damage mechanisms in a selection of notched composite structures under multiaxial fatigue loading. Thanks to a synergistic combination of X-ray micro-computed tomography ($\mu$-CT) and Digital Image Correlation (DIC), the main failure modes are identified while the crack volume associated to each mechanism is characterized. This study provides unprecedented quantitative data for the development and validation of computational models to capture the fatigue behavior of polymer composite structures.

\end{abstract}


\end{frontmatter}


\section{Introduction}

The broad application of polymer composites in engineering requires the insightful understanding of their damage mechanisms both under quasi-static and fatigue loading. These mechanisms affect the mechanical behavior of composites significantly \cite{sal2,patent,sal1,kirane,kirane2} and lead to important size effects that must be carefully considered in structural design \cite{Baz1,Sal,Yao1,Yao2,Yao4,Seung1,Seung2,Deleo,sal5}.

Towards this important goal, a range of non-destructive inspection techniques has been developed and applied in the last few decades. The most attractive techniques include acoustic emission \cite{AE,AE2,AE3}, infrared thermography \cite{thermo,thermo2,thermo3,thermo4}, ultrasonic C-scan \cite{Cscan}, Digital Imaging Correlation \cite{digital,digital2} and X-ray micro-computed tomography \cite{CT,CT2,CT3}. An analysis to combine the mechanical behavior of the materials with the damage evaluation via the forgoing useful tools promotes a better understanding on the fracturing behavior of composite structures under a variety of loading conditions. This is a condition of utmost importance for the development of safe and reliable designs and certification guidelines.

Among these techniques, acoustic emission method is widely used for monitoring damage initiation. An excellent investigation on the initiation of transverse cracking and delamination in textile composite under uni-axial tensile test was provided by Lomov \emph{et al.} \cite{AE3}. Similar studies were reported on the unidirectional and woven composites by Silversides \emph{et al.} \cite{AE2}.

However, infrared thermography and ultrasonic C-scan are typically employed to have a rough estimation of the damage progression in composite structures. A noticeable example was provided by Meola and Carlomagno \cite{thermo2} who evaluated impact damage in Glass Fiber Reinforced Polymer (GFRP) leveraging the thermographic technique. Thanks to this method, the impact damage progression throughout the thickness of the laminates was successfully reconstructed. By conducting an ultrasonic inspection on the damaged specimens, Scarponi and Briotti \cite{Cscan} identified significant development of delamination in thermoset polymer reinforced by various types of fibers.  

Although the forgoing techniques led to a remarkable progress in the damage characterization, the identification of individual damage types is still not clear since the damage analyzed using the forgoing inspection tools is usually homogenized. In this context, microscopic techniques (Optical and Scanning Electron Microscopy) are extensively used to visualize different damage types in specimens. Carraro \emph{et al.} \cite{carraro}, for instance, conducted an optical microscopic inspection which was capable of capturing the delamination induced by transverse cracking in cross-ply laminates under tensile fatigue loading. On similar grounds, a micro-damage study leveraging Scanning Electron Microscopy (SEM) was provided by Nguyen \emph{et al.} \cite{minh} who identified shear hackles between the fibers in specimens featuring [+45/$-$45]$_{4s}$ layup under un-axial tensile test. However, these techniques can only provide a local visualization of a detailed damage morphology since the specimens must be cut into a small portion which can be properly operated by microscopic techniques.

Thanks to the emergence and development of X-ray micro-computed tomography ($\mu$-CT), a detailed damage morphology throughout the entire composite structure can be obtained even under an in-situ inspection test. An interesting study was provided by Yu \emph{et al.} \cite{CT2} who investigated the damage evolution in 3D woven composite under tensile fatigue loading. It was concluded that the dominating damage mechanism before final failure is the debonding cracks spreading along the interface between binder yarns and matrix. By leveraging $\mu$-CT technique, the growth of different type of damage was also successfully quantified as a function of fatigue lifetime.

A further step towards an even more insightful damage analysis requires a synergistic approach combing $\mu$-CT technique and other aforementioned methods. In this way, this work presents a comprehensive damage analysis via X-ray micro-computed tomography and Digital Imaging Correlation techniques on notched quasi-isotropic [+45/90/$-$45/0]$_{s}$ and cross-ply [0/90]$_{2s}$ laminates under multi-axial quasi-static and fatigue loading. Thanks to the synergistic approach, a detailed analysis of the damage mechanisms and quantification of damage evolution was achieved. This result, which is of utmost importance for the development and validation of fatigue computational models for notched composite structures under multiaxial loading, has never been obtained before.

\section{Materials and Methods} 
\subsection{Material Preparation}
Following \cite{Yaomulti}, quasi-isotropic [+45/90/$-$45/0]$_{s}$ and cross-ply [0/90]$_{2s}$ laminates made of a Glass Fiber Reinforced Polymer (GFRP) by Mitsubishi Composites \cite{Rock} were investigated in this work. The material system included 7781 unidirectional E-glass fiber and NB301 epoxy resin leading to a 0.28 mm lamina with 68\% fiber volume fraction. The glass transition temperature for epoxy resin is about $120^{\circ}$C. The laminates were manufactured from prepreg sheets which were hand laid-up and then vacuum-bagged using a Vacmobile mobile vacuum system \cite{pump}. A Despatch LAC1-38A programmable oven was used to cure the panels by ramping up the temperature from room temperature to $135^{\circ}$C in one hour, soaking for one hour, and cooling down to room temperature. 

\subsection{Test Method} %{\color{blue} what was R? Finish, YQ}

The application of the combinations of nominal normal and shear stresses on notched laminates was achieved by using an Arcan rig (Fig. \ref{fig:procedure}a). The multi-axial quasi-static and fatigue tests were performed in a servo-hydraulic 8801 Instron machine with closed-loop control. The load control with a stress ratio of $R=0.1$ and a low frequency of $f=5$ Hz was conducted for the fatigue tests. To describe the multi-axial loading configuration, multiaxiality ratio can be defined as $\lambda=$ arctan$(\tau_{N}/\sigma_{N})$ where $\sigma_{N} = P$cos$\theta/[(w-a_0)t]$ is the nominal normal stress and $\tau_{N} = P$sin$\theta/[(w-a_0)t]$ is the nominal shear stress applied to the specimen. In the definition of stress, $P$ is the instantaneous load, $w$ is the specimen width, $a_0$ is the crack length or hole diameter, $t$ is the specimen thickness and $\theta$ is the angle between loading direction and axial direction of the specimen. A detailed description on the test method and loading conditions can be found in \cite{Yaomulti}.

\subsection{Specimen preparation}
%{\color{blue} where is $\lambda$ defined? Moved Test Method section before specimen preparation, YQ. What was the acquisition frequency for DIC? Finished, YQ}

\begin{figure} [H]
\center
\includegraphics[scale=0.37]{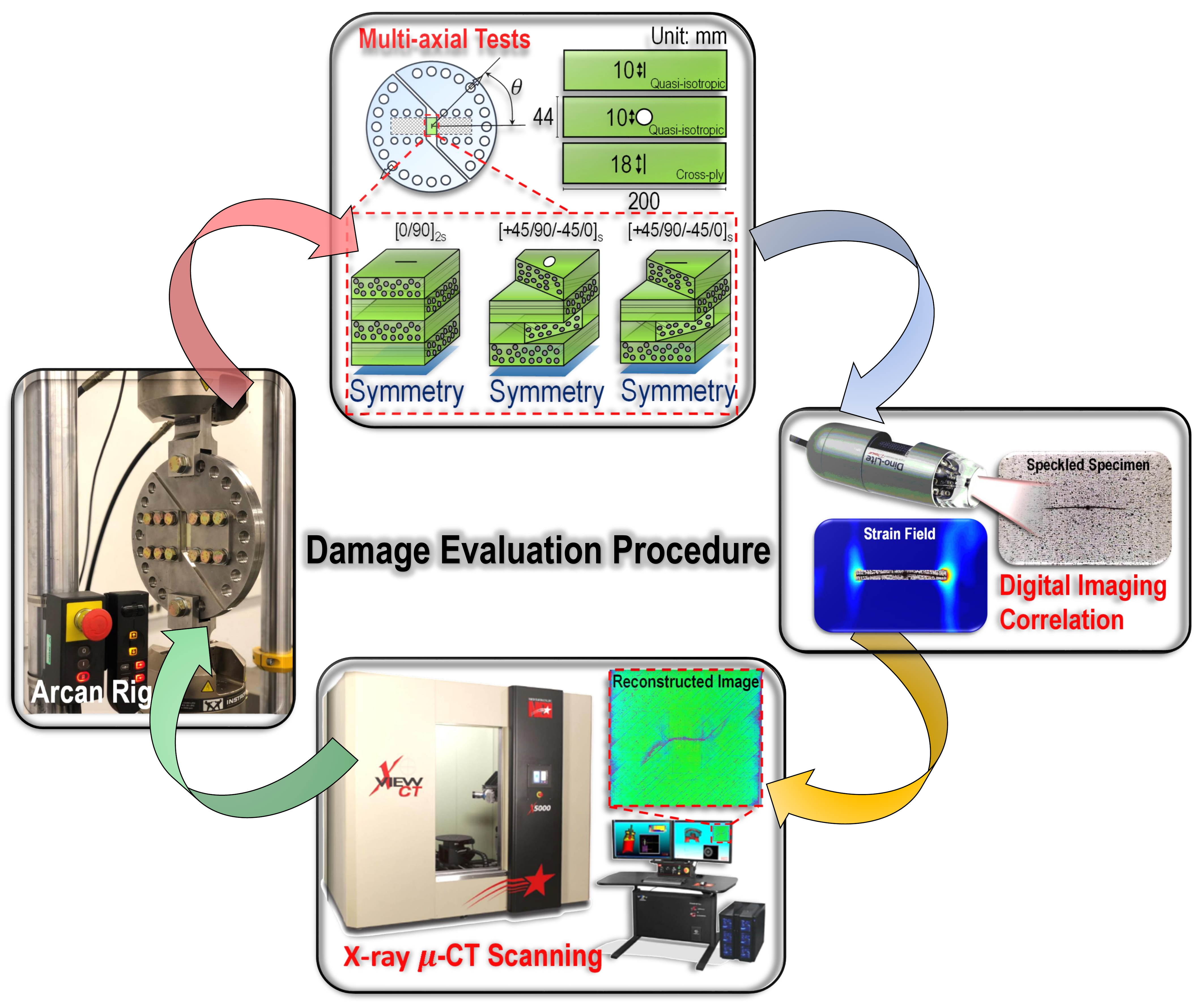}
\caption{Notched specimen geometries and damage evaluation leveraging Digital Imaging Correlation (DIC) and X-ray micro-computed tomography ($\mu$-CT).}
\label{fig:procedure}
\end{figure}

To study the damage evolution of notched laminates under multi-axial tests, specimens with the same dimensions but featuring different types of notches were prepared as shown in Figure \ref{fig:procedure}. The detailed description of the manufacturing of these notches can be found in the previous work by Qiao \emph{et al.} \cite{Yaomulti}. The procedure of damage evaluation leveraging Digital Image Correlation (DIC) and X-ray micro-computed tomography ($\mu$-CT) is illustrated in Figure \ref{fig:procedure} and described in the following sections.

\subsubsection{Digital Image Correlation (DIC)}
\label{sec:specimenpreparation}
The strain distribution of notched laminates was investigated for three multiaxiality ratios ($\lambda=0, 0.785$ and $1.571$) by using an open source DIC system programmed in MATLAB software developed at the Georgia Institute of Technology  \cite{Blader,harilal}. To this end, a recording of videos with a frame rate of 30 fps for speckled specimens was analyzed at roughly 97\% of total quasi-static or fatigue life. The total quasi-static life corresponded to its final catastrophic failure whereas the percentage of fatigue life corresponded to its performance at the peak load of applied cyclic load (55\% of quasi-static critical load). 

\subsubsection{X-ray micro-computed tomography ($\mu$-CT)}
\label{sec:specimenpreparation2}
The sub-critical damage of notched laminates was studied in detail by means of a NSI X5000 X-ray micro-tomography scanning system \cite{northstar} with a X-ray tube setting of 90 kV in voltage and 220 $\mu$A in current. To have a better understanding of the damage evolution under multi-axial tests, damaged specimens taken from the multi-axial quasi-static tests were analyzed at 90\% and even later stage of the entire life whereas the ones taken from the multi-axial fatigue tests were analyzed at 40\% and 70\% of the entire life. It is worth mentioning here that the total quasi-static life corresponded to its life reaching the critical load whereas the total fatigue life corresponded to its lifetime to failure.

Prior to the scanning, a dye penetrant composed of zinc iodide powder (250 g), isopropyl alcohol (80 ml), Kodak photo-flow solution (1 ml) and distilled water (80 ml) was used as a supplement to improve the visualization of the damage mechanisms \cite{dye,dye1}. The specimens were soaked in the dye penetrant mixture for approximately one day. The sub-critical damage in each ply and interface were identified by slicing through the reconstructed 3D images of the specimens via the software ImageJ \cite{imagej}.

\section{Digital Image Correlation (DIC) Analysis}
\label{sec:DIC}
The sub-critical, distributed damage occurring under fatigue loading of notched composite structures generally leads to a local reduction of the material stiffness with consequent increase of the local strain. Accordingly, to have a qualitative idea of the overall damage distribution notched [0/90]$_{2s}$ and [+45/90/$-$45/0]$_{s}$ specimens were analyzed through DIC as described in section \ref{sec:specimenpreparation}. 

\begin{figure} [H]
\center
\includegraphics[scale=0.19]{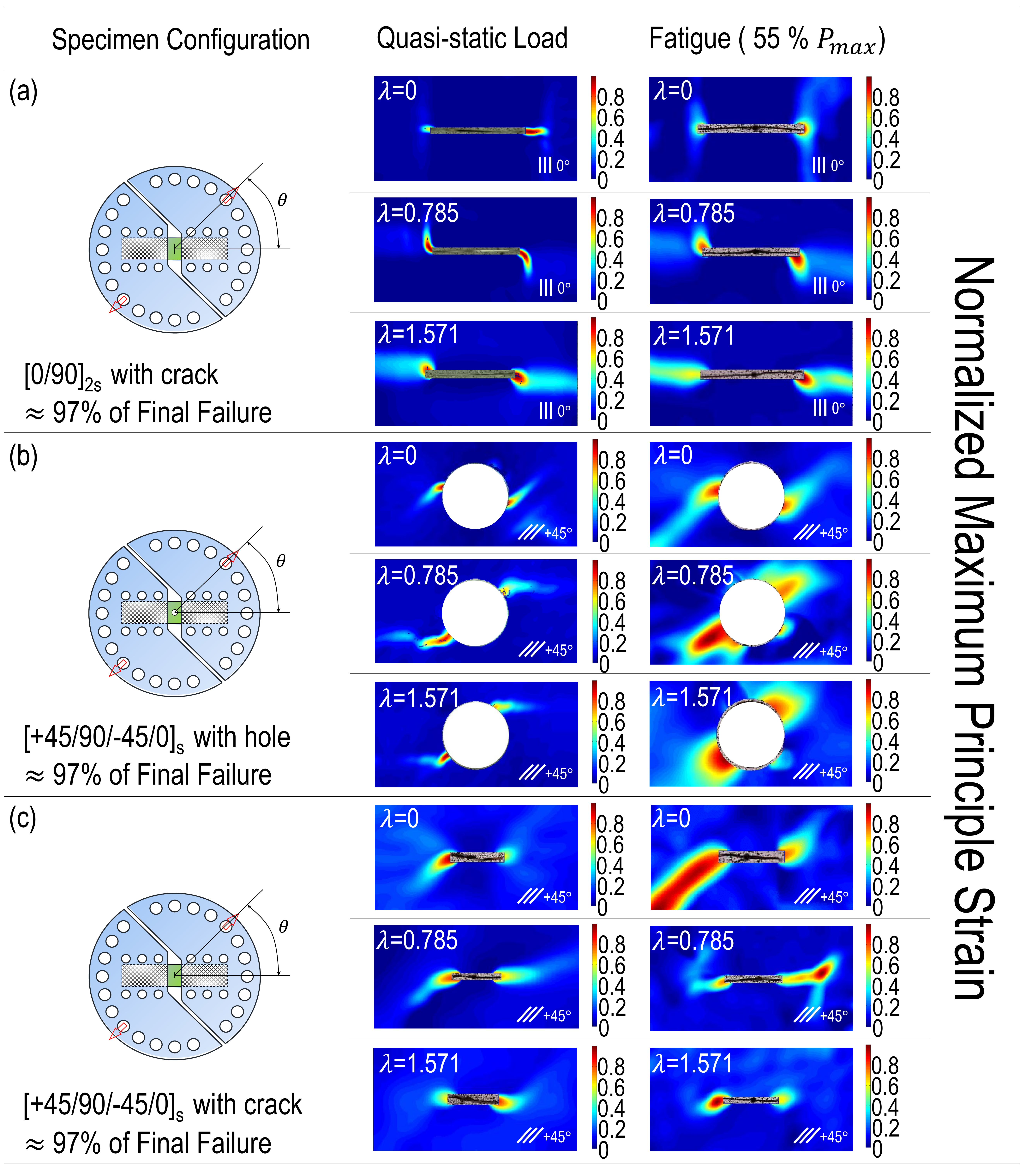}
\caption{Digital Imaging Correlation (DIC) analysis of investigated notched specimens at 97\% of total quasi-static or fatigue life. This figure compares three multiaxiality ratios. From the contour plots of the normalized maximum principle strain, it can be noted that the region of high deformation in fatigue is generally more spread compared to quasi-static loading. This is an indication that fatigue damage is generally more distributed than its quasi-static counterpart.}
\label{fig:DIC}
\end{figure}

As illustrated in Figure \ref{fig:DIC} which compares the maximum principal strain distribution of the quasi-static and fatigue loading, the overall damage distribution under fatigue loading appears to be more diffuse compared to the quasi-static case. A similar phenomenon was found by Fujii \emph{et al.} \cite{fujii1,fujii2,fujii3} who tested tabular specimens featuring a circular hole under the multi-axial fatigue loading of tension and torsion. This result is an indication of a larger non-linear Fracture Process Zone (FPZ) in the specimens subject to fatigue loading. It is worth mentioning that the maximum principle strains were normalized against their maximum values in the area of interest in order to have a comparison between quasi-static and fatigue loading condition. In fact, the focus was not on the magnitude of the strain field but the overall damage distribution.

\begin{figure} [H]
\center
\includegraphics[scale=0.78]{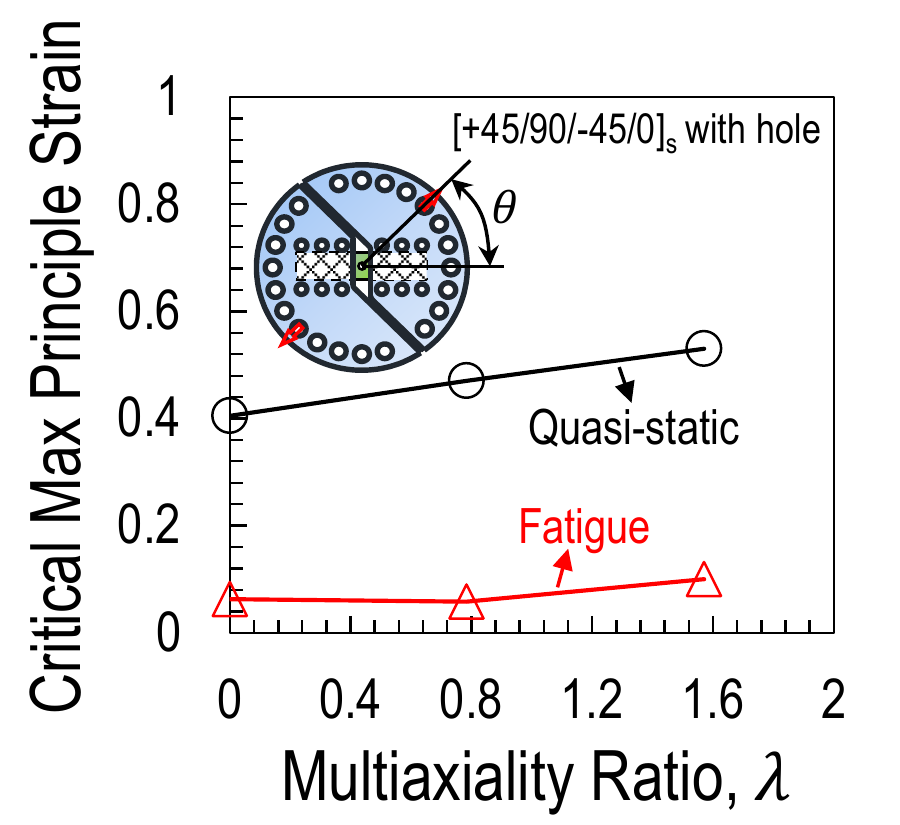}
\caption{The evolution of critical maximum principle strain at 97\% of final failure for the specimen weakened by a central hole as a function of multiaxiality ratio. The graph compares quasi-static and fatigue results.}
\label{fig:DICstrain}
\end{figure}

However, the critical magnitude in the maximum principle strain field for quasi-isotropic specimens featuring a central hole under multi-axial tests provides an insightful information on the quasi-static and fatigue fracturing behavior. As can be noted in Figure \ref{fig:DICstrain}, the critical magnitude of the maximum principle strain in fatigue case is significantly lower than the one in quasi-static case almost before catastrophic failure. This result is an indication of substantially different failure criteria for quasi-static and fatigue case which is typically reported in fatigue fracturing behavior of thermoset polymers showing that the Mode I critical stress intensity factor in fatigue case is generally lower than the one in quasi-static case \cite{suresh,geubelle, guo}.

\begin{figure} [H]
\center
\includegraphics[scale=0.7]{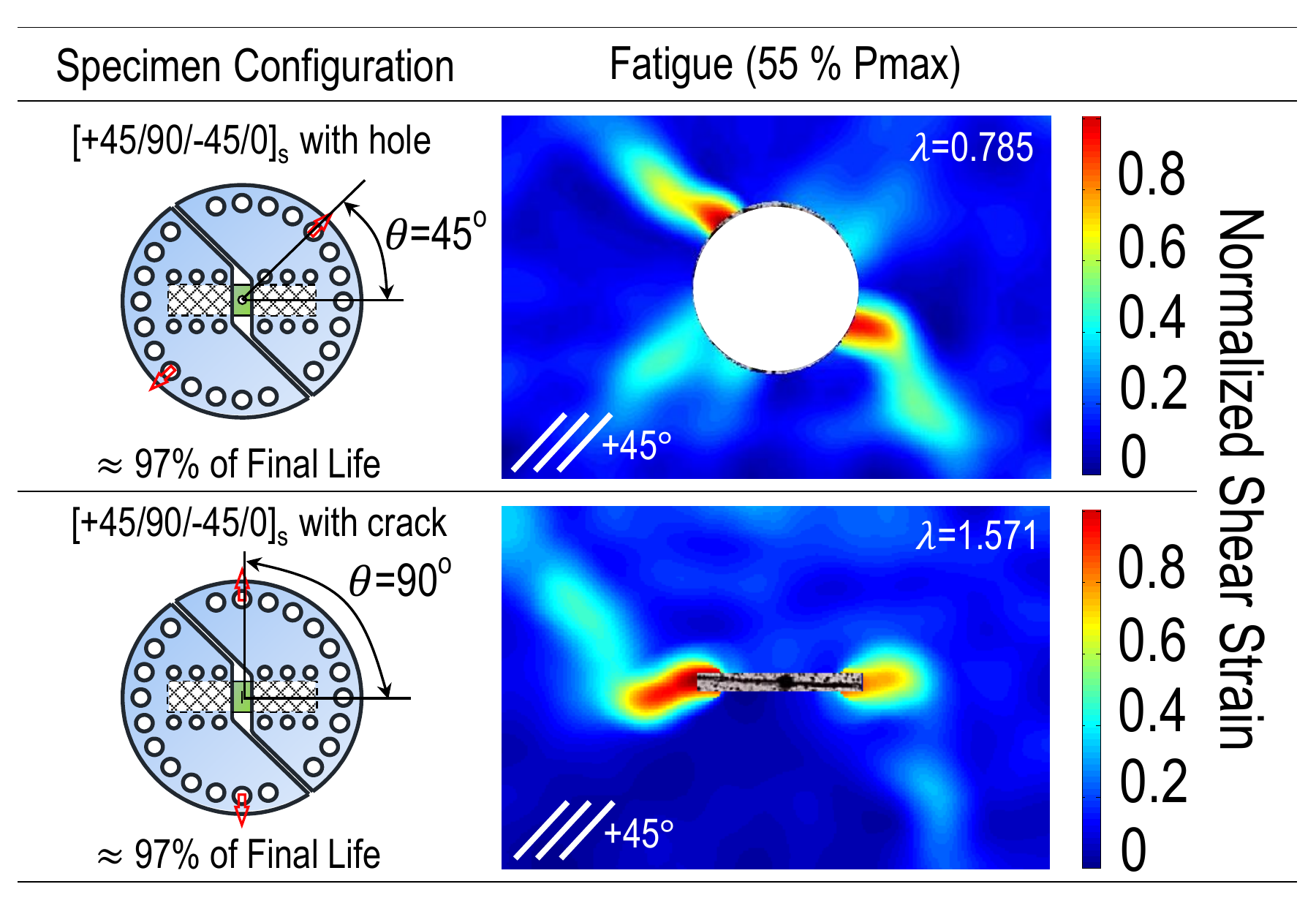}
\caption{Digital Imaging Correlation (DIC) analysis of notched quasi-isotropic specimens for multiaxiality ratio $\lambda=0.785$ and $\lambda=1.571$ at 97\% of total fatigue life. Note that the shear strain was normalized against its maximum value in the area of interest.}
\label{fig:DICshear}
\end{figure}

On the other hand, before moving to the following section on the investigation of damage mechanisms via $\mu$-CT technique, the fracturing features can be roughly known based on the overall strain distribution leveraging DIC. As can be noted in Figure \ref{fig:DIC}, the region of high deformation for notched laminates at various multiaxiality ratios does not show significant differences in terms of shape and location for quasi-static and fatigue case. For notched cross-ply laminates as shown in Figure  \ref{fig:DIC}a, the specimen under pure tension is characterized by a region of high deformation in $0^{\circ}$ plies at the notch tip whereas the one under pure shear features a noticeable damage band in $90^{\circ}$ plies ahead of the notch. This corresponds to splitting and fiber kinking for pure tension and pure shear case respectively. For notched quasi-isotropic laminates as shown in Figures \ref{fig:DIC}b-c, the specimens at various multiaxiality ratios are mainly characterized by a highly-deformed region in $+45^{\circ}$ plies in front of the notch which is associated to significant splitting. In addition to this, a highly-deformed region due to shear strain almost perpendicular to the $+45^{\circ}$ plies was observed for the specimens under shear-dominated loading as shown in Figure \ref{fig:DICshear}. This indicates that the $+45^{\circ}$ plies are under compression leading to the potential micro-buckling of the fibers.

% More interestingly, the quasi-isotropic specimen in presence of an open hole or central crack under shear-dominated fatigue loading  exhibits different fracturing features compared to the one under quasi-static loading. By taking an example of the case $\lambda=0.785$, as it can be noted from Figure \ref{fig:DICfailure}, a striking feature of straight micro-buckling paths near the hole characterizes the specimen under quasi-static loading despite the final failure due to the micro-buckling almost perpendicular to the +$45^{\circ}$ plies whereas less pronounced straight micro-buckling exhibits in the fatigue case and the dominant path propagates almost perpendicular to the +$45^{\circ}$ plies. The detailed discussion will be presented in the next section.
 %{\color{blue} This part needs to be explained better. Also the figure needs to be improved. As of now, the location of the microbuckling is not clear. Also, the figure does not show very well the difference between one mechanism and the other. MS}

\section{Quantitative analysis of the damage mechanisms}
\label{sec:damagemechanism}
\begin{figure} [H]
\center
\includegraphics[scale=0.8]{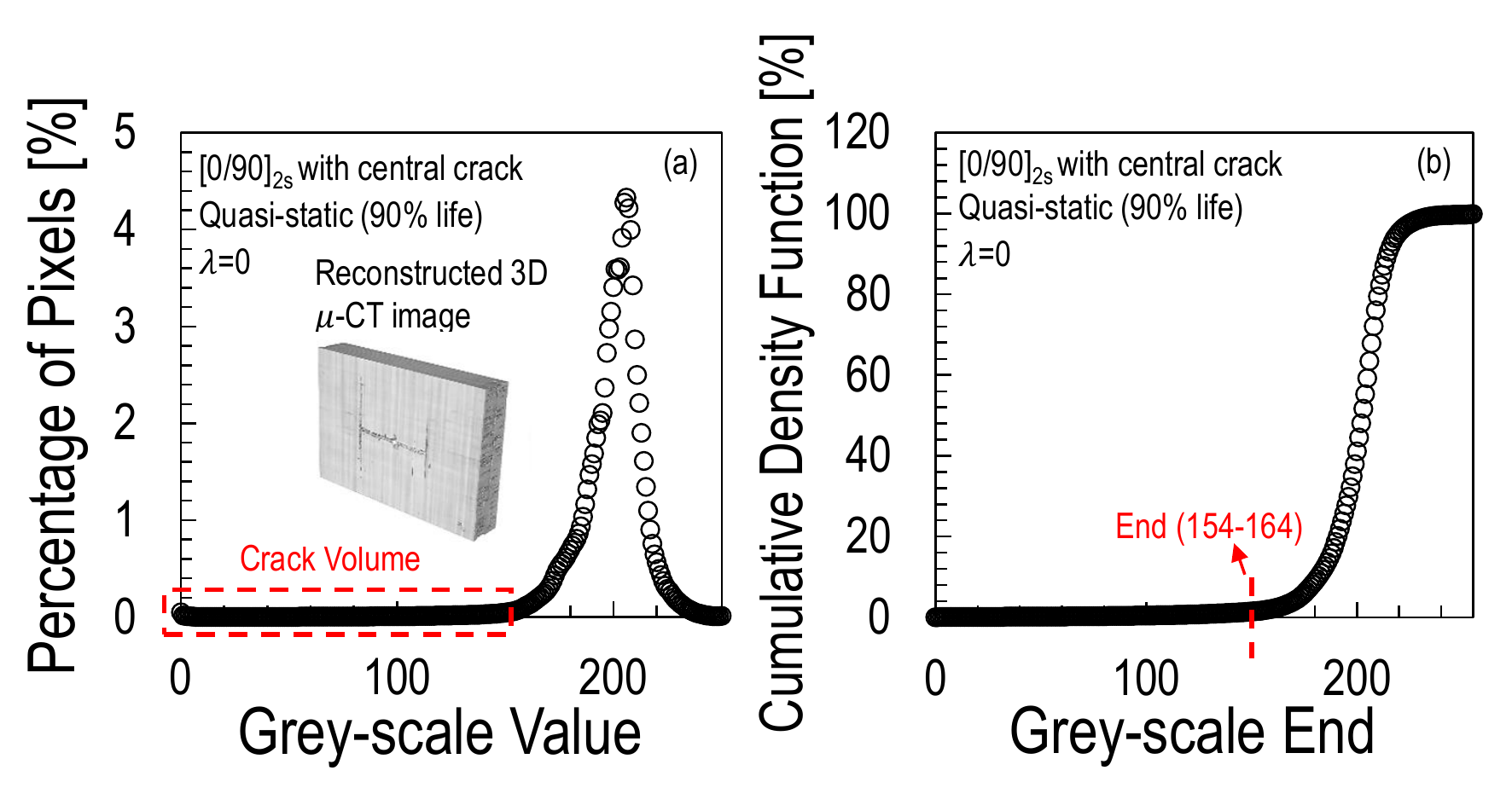}
\caption{(a) Typical graph of the percentage of pixels vs. grey-scale value in the gauge volume of the $[0/90]_{2s}$ specimens weakened by a 18 mm central crack for a multiaxiality ratio $\lambda=0$ based on the reconstructed 3D image obtained by micro-computed tomography; (b) The evolution of cumulative density function of grey-scale distribution as a function of grey-scale end. All the grey-scale values lower than this end was considered as the damage. In this case, the grey-scale end (159) was selected and $\pm$5 was considered to account for the measurement errors on the accuracy of this selective end value.}
\label{fig:pixels}
\end{figure}

To shed more light on the characteristics of the fracturing morphology of notched laminates under multi-axial quasi-static and fatigue loading, damaged specimens were analyzed leveraging X-ray $\mu$-CT using the parameters described in the section \ref{sec:specimenpreparation2}. This non-destructive technique was particularly important to guarantee that no additional damage was created during the damage visualization process. Thanks to this technique, the sub-critical damage can be observed through the reconstructed 3D images in order to have a quantitative comparison on the crack volume and delamination area of the specimens under multi-axial tests. With the supplement of the dye penetrant, the sub-critical damage can be easily visualized as illustrated in Figures \ref{fig:damagemechanismcrossply0}-\ref{fig:damagemechanismquasiisocrack90}. Thanks to the great contrast provided by the use of the dye penetrant, the analysis of the grey-scale value can be used to roughly estimate the crack volume and delamination area of the specimen. Figure \ref{fig:pixels}a shows an example of such analysis for a cross-ply specimen featuring a central crack under the quasi-static loading in pure tension. In this plot, the lower grey-scale values represent the sub-critical damage in the gauge volume of the specimen while the higher grey-scale values, constituting the most part of the percentage of pixels, characterize the remaining part of the gauge volume of the specimen. It is worth mentioning here that there are roughly 90 reconstructed images throughout the thickness of the specimen leading to about 2.5 $\times$ $10^{7}$ pixels for the gauge volume of the specimen. To quantify the sub-critical damage, as illustrated in Figure \ref{fig:pixels}b, the cumulative density function of grey-scale distribution up to a range of grey-scale end was treated as crack volume to provide a safe estimation. This range was based on a reference value with the upper and lower bounds of 5 grey-scale values while this reference value was about 144 to 179 for different specimens and reasonably taken on the location of the specimen where no significant damage was observed.

%{\color{blue} Also, I think you should give the size of your pixels. If based on the size of your volume you can estimate it. This way, if one wants to reconstruct the size of the crack volume, he can do it. MS}

\subsection{$[0/90]_{2s}$ laminate with a central crack}
\subsubsection{Fatigue loading condition}
\begin{figure} [H]
\center
\includegraphics[scale=0.175]{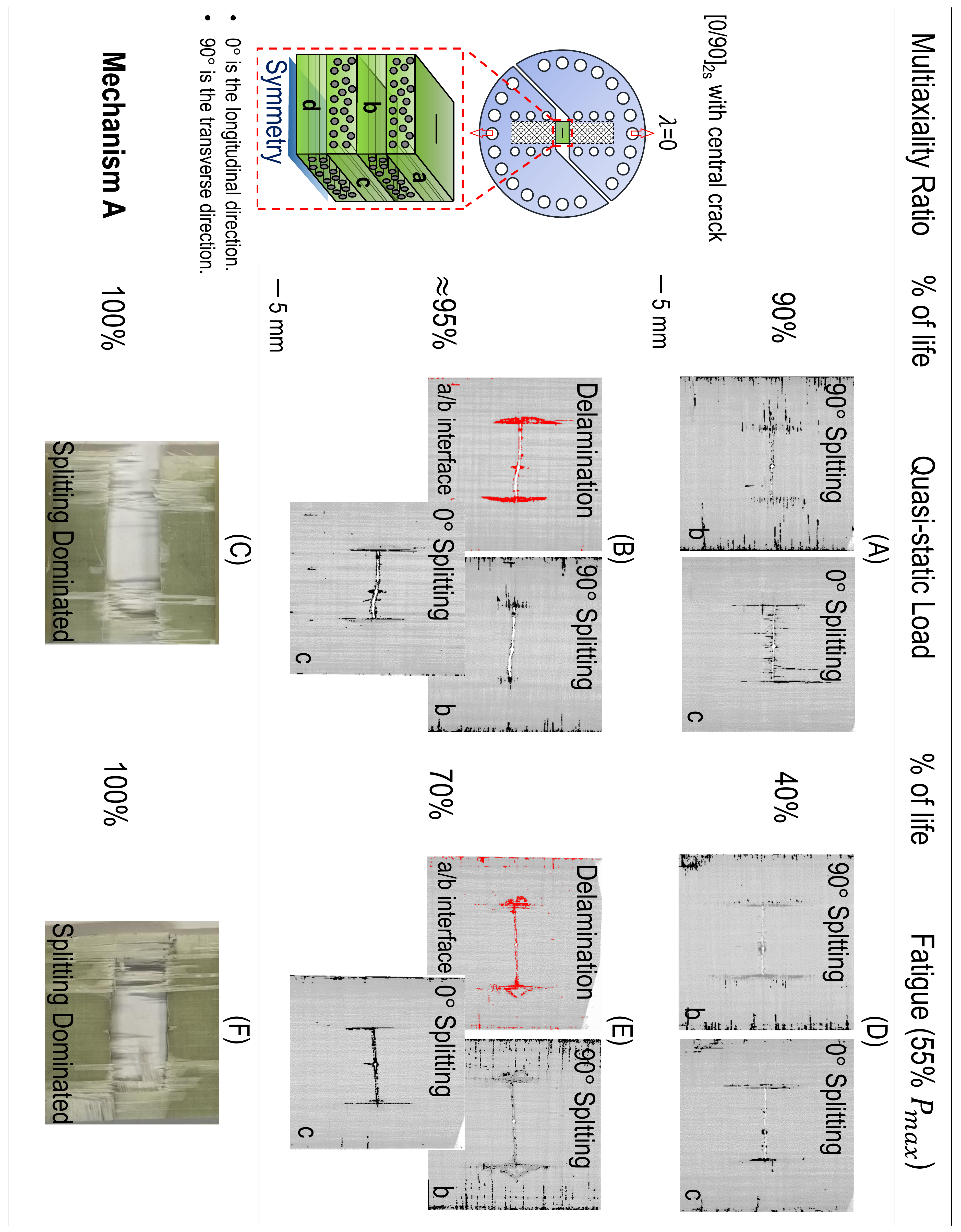}
\caption{Analysis of quasi-static and fatigue damage mechanisms by micro-computed tomography. Comparison between quasi-static and fatigue damage in $[0/90]_{2s}$ specimens weakened by a 18 mm central crack for a multiaxiality ratio $\lambda=0$ in the $0^{\circ}$,  $90^{\circ}$ layers and at the interface. Note that the original colors of the images were inverted for a better visualization on the damage.}
\label{fig:damagemechanismcrossply0}
\end{figure}

\leftline{(a) Mechanism A}
In the case of specimens subjected to tension-dominated loading conditions ($\lambda=0, 0.262$), the dominant mechanisms before sudden failure are the splitting in $0^{\circ}$ plies at the notch tip and the splitting in $90^{\circ}$ plies. The final failure is triggered by the additional splitting in $0^{\circ}$ plies away from the notch tip and the breakage of $0^{\circ}$ fibers. As illustrated in Figure \ref{fig:damagemechanismcrossply0}, the damage evolution as a function of fatigue lifetime for mechanism A can be summarized as follows: \vspace{0.4cm}

(1) splitting initiates in $90^{\circ}$ plies (Fig. \ref{fig:damagemechanismcrossply0}D); 

(2) splitting initiates in $0^{\circ}$ plies at notch tip (Fig. \ref{fig:damagemechanismcrossply0}D); 

(3) splitting develops in the corresponding piles (Fig. \ref{fig:damagemechanismcrossply0}E);

(4) a small amount of delamination between $0^{\circ}$ and $90^{\circ}$ plies starts to initiate (Fig. \ref{fig:damagemechanismcrossply0}E);

(5) additional splitting in $0^{\circ}$ plies away from the notch tip happens simultaneously 

with the breakage of $0^{\circ}$ fibers (Fig. \ref{fig:damagemechanismcrossply0}F). \vspace{0.4cm}

\begin{figure} [H]
\center
\includegraphics[scale=0.175]{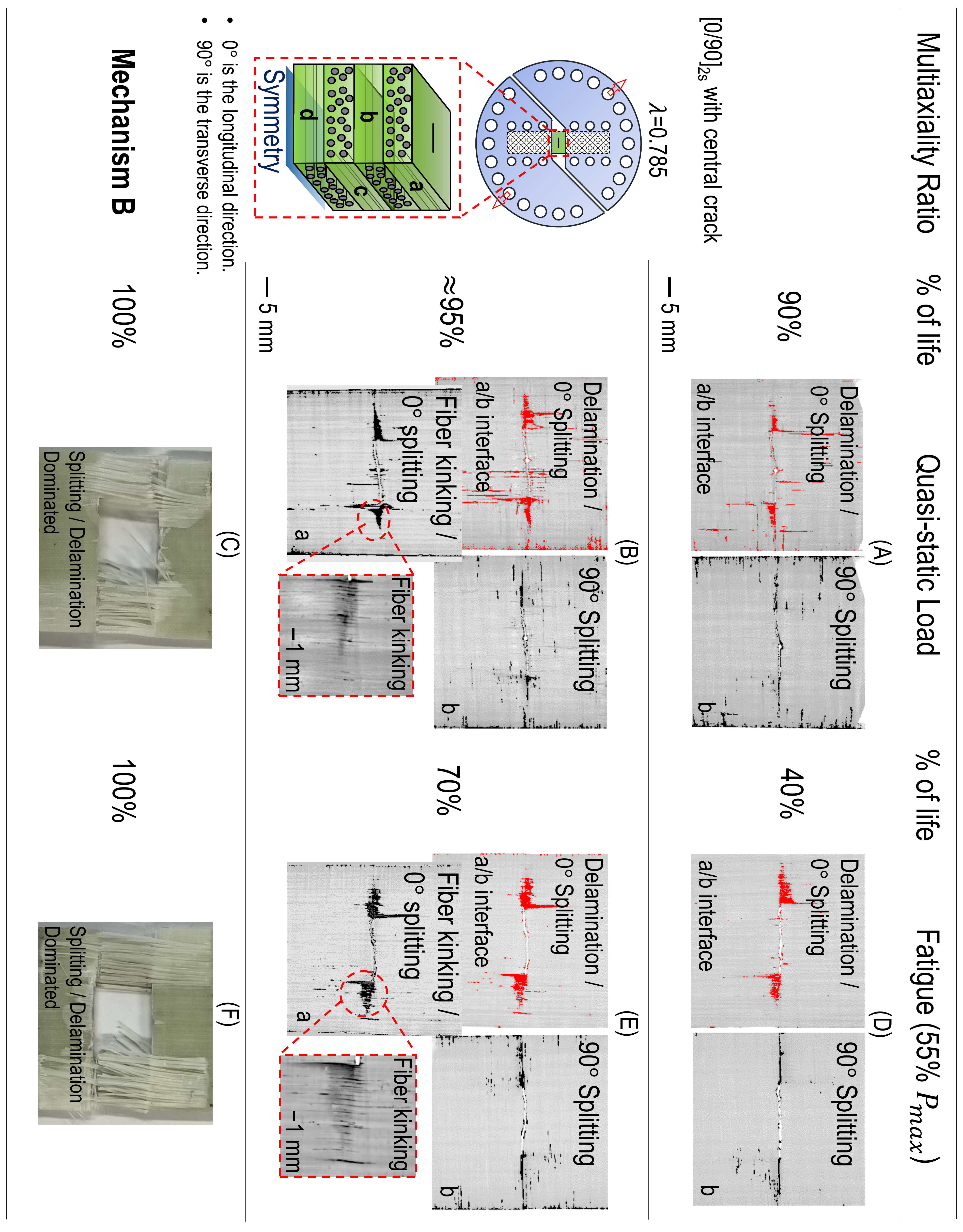}
\caption{Analysis of quasi-static and fatigue damage mechanisms by micro-computed tomography. Comparison between quasi-static and fatigue damage in $[0/90]_{2s}$ specimens weakened by a 18 mm central crack for a multiaxiality ratio $\lambda=0.785$ in the $0^{\circ}$,  $90^{\circ}$ layers and at the interface. Note that the original colors of the images were inverted for a better visualization on the damage.}
\label{fig:damagemechanismcrossply45}
\end{figure}

\leftline{(b) Mechanism B}
When the multiaxiality ratio $\lambda=0.785, 1.309$, a mix of nominal normal and shear stresses is applied on specimens and failure behavior follows Mechanism B. In this case, the dominant mechanism transitions from splitting to a combination of delamination and splitting. The final failure happens with the growth of splitting in $0^{\circ}$ plies at the notch tip and the delamination between $0^{\circ}$ and $90^{\circ}$ plies. As illustrated in Figure \ref{fig:damagemechanismcrossply45}, the damage evolution of mechanism B can be summarized by the following steps: \vspace{0.4cm}

(1) splitting initiates\ in $90^{\circ}$ plies (Fig. \ref{fig:damagemechanismcrossply45}D); 

(2) splitting initiates in $0^{\circ}$ plies at notch tip simultaneously with the 

formation of delamination between $0^{\circ}$ and $90^{\circ}$ plies (Fig. \ref{fig:damagemechanismcrossply45}D);

(3) splitting and delamination develop in/between the corresponding plies (Fig. \ref{fig:damagemechanismcrossply45}E);

(4) fibers in $0^{\circ}$ plies start to kink due to the shear stress (Fig. \ref{fig:damagemechanismcrossply45}E);

(5) delamination between $0^{\circ}$ and $90^{\circ}$ plies grows together with the significant 

splitting at notch tip (Fig. \ref{fig:damagemechanismcrossply45}E-F).\vspace{0.4cm}

\begin{figure} [H]
\center
\includegraphics[scale=0.165]{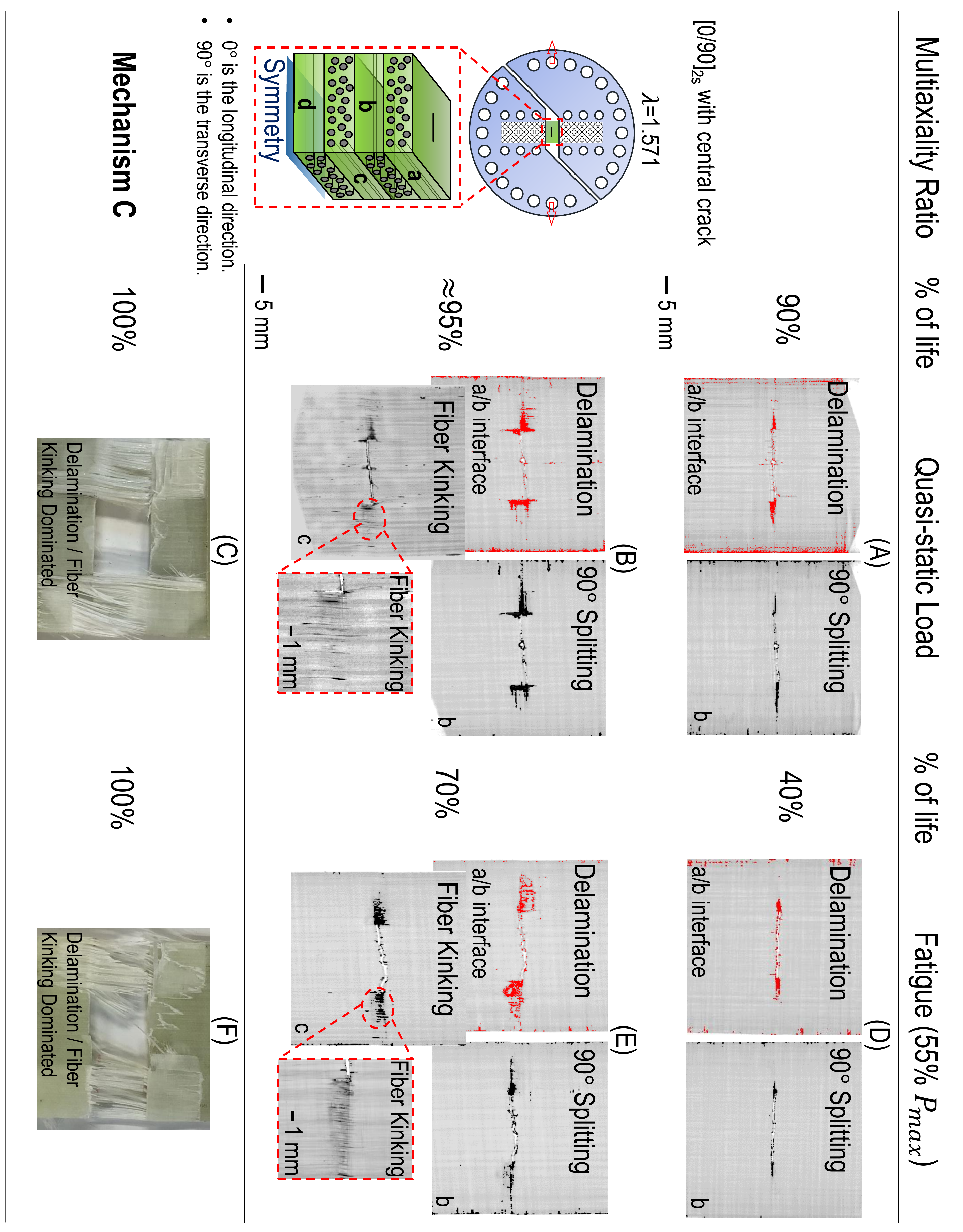}
\caption{Analysis of quasi-static and fatigue damage mechanisms by micro-computed tomography. Comparison between quasi-static and fatigue damage in $[0/90]_{2s}$ specimens weakened by a 18 mm central crack for a multiaxiality ratio $\lambda=1.571$ in the $0^{\circ}$,  $90^{\circ}$ layers and at the interface. Note that the original colors of the images were inverted for a better visualization on the damage.}
\label{fig:damagemechanismcrossply90}
\end{figure}

\leftline{(c) Mechanism C}
When the specimens are only subjected to pure shear loading, the dominant mechanisms are the delamination between $0^{\circ}$ and $90^{\circ}$ plies and then fiber kinking in $0^{\circ}$ plies. The final failure is due to the unstable growth of the inter-laminar crack enabling the splitting in $0^{\circ}$ plies at notch tip. As illustrated in Figure \ref{fig:damagemechanismcrossply90}, the damage evolution of mechanism C consists of the following phases: \vspace{0.4cm}

(1) a small amount of splitting initiates in $90^{\circ}$ plies at notch tip (Fig. \ref{fig:damagemechanismcrossply90}D); 

(2) delamination initiates between $0^{\circ}$ and $90^{\circ}$ plies (Fig. \ref{fig:damagemechanismcrossply90}D);

(3) fibers in $0^{\circ}$ plies start to kink significantly due to the shear stress (Fig. \ref{fig:damagemechanismcrossply90}E);

(4) delamination develops between the corresponding plies (Fig. \ref{fig:damagemechanismcrossply90}E);

(5) delamination grows dramatically and drives the significant splitting in $0^{\circ}$ plies 

at notch tip (Fig. \ref{fig:damagemechanismcrossply90}E-F).

%{\color{blue} Maybe you can add labels (a), (b) etc in the figures and you can refer every step to the various labels when you describe the damage evolution of Mechs A,B, and C. MS}

\subsubsection{Quantitative analysis: fatigue vs. quasi-static loading}
\label{sec:quantitativecrossply}
\begin{figure} [H]
\center
\includegraphics[scale=0.28]{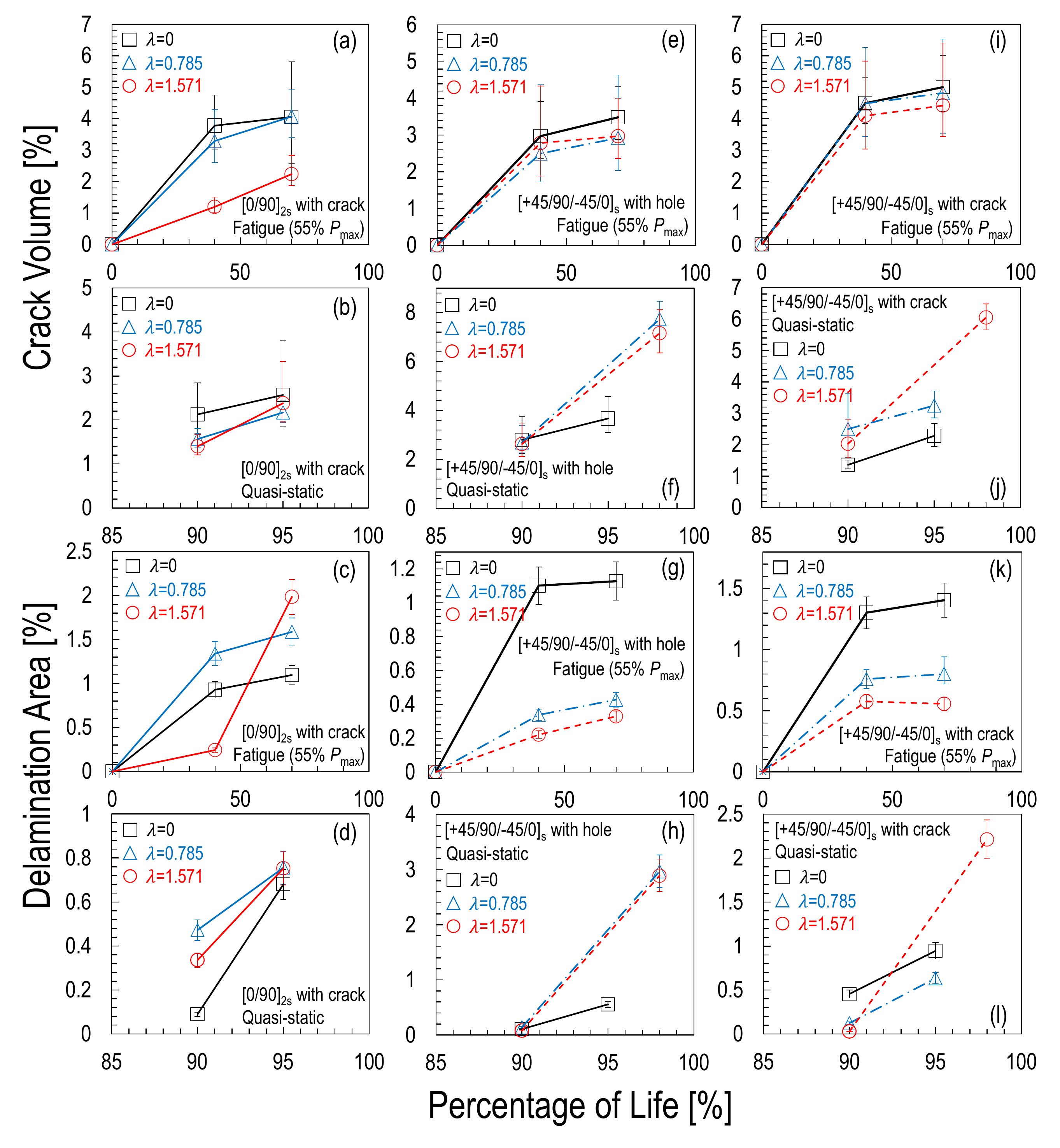}
\caption{The evolution of the total crack volume and delamination area for all the investigated specimens weakened by a central crack or hole as a function of the percentage of life for three multiaxiality ratios. The graphs compare quasi-static and fatigue results.}
\label{fig:damageevolution}
\end{figure}

The crack volume and delamination area as a function of the percentage life were plotted in Figures \ref{fig:damageevolution}a-d for specimens under quasi-static and fatigue loading. As illustrated in Figure \ref{fig:damageevolution}c, in the fatigue case, the specimens following mechanism B and C feature larger delamination area compared to the specimens following mechanism A at 70\% of total fatigue life, with 1.6\% for $\lambda=0.785$ and 2\% for $\lambda=1.571$ but only 1\% for $\lambda=0$. This is an indirect evidence of the foregoing damage mechanisms showing that delamination contributes to most of the energy dissipation in notched cross-ply laminates under shear-dominated loading. However, this is not the case for total crack volume in specimens at the corresponding life as shown in Figure \ref{fig:damageevolution}a, with the lowest value (2.2\%) featuring $\lambda=1.571$ but 4.1\% for $\lambda=0$ and $0.785$. This lower crack volume explains the less pronounced stiffness degradation of specimens under the fatigue of pure shear as shown in \cite{Yaomulti} and is associated to the significant reduction of splitting in $90^{\circ}$ plies.

With the comparison to the fatigue case, similar damage mechanisms were observed in the case of quasi-static loading for all the multiaxiality ratios as shown in Figures \ref{fig:damagemechanismcrossply0}-\ref{fig:damagemechanismcrossply90}. However, the overall sub-critical damage during quasi-static loading is less diffused, consistent with the DIC results discussed in section \ref{sec:DIC} and the quantitative analysis via $\mu-$CT. In Figures \ref{fig:damageevolution}a-d, for all the multiaxiality ratios, both crack volume and delamination area for the quasi-static case above 90\% of total quasi-static life are significantly less than the fatigue case at even lower percentage of the life. Only about  2.4\% crack volume and 0.7\% delamination area in average for different multiaxiality ratios were observed at 95\% of total quasi-static life. Notwithstanding this, these sub-critical damage grows rapidly close to the end of total quasi-static life leading to the catastrophic failure of the specimen. It is worth mentioning here that the specimen only at 90\% and even later stage of its quasi-static life was prepared for the quantitative analysis due to the less pronounced damage at the early stage of the quasi-static loading and the difficulties for the damage visualization.

\subsection{$[+45/90/-45/0]_s$ laminate with an open hole}
\subsubsection{Fatigue loading condition}
\begin{figure} [H]
\center
\includegraphics[scale=0.14]{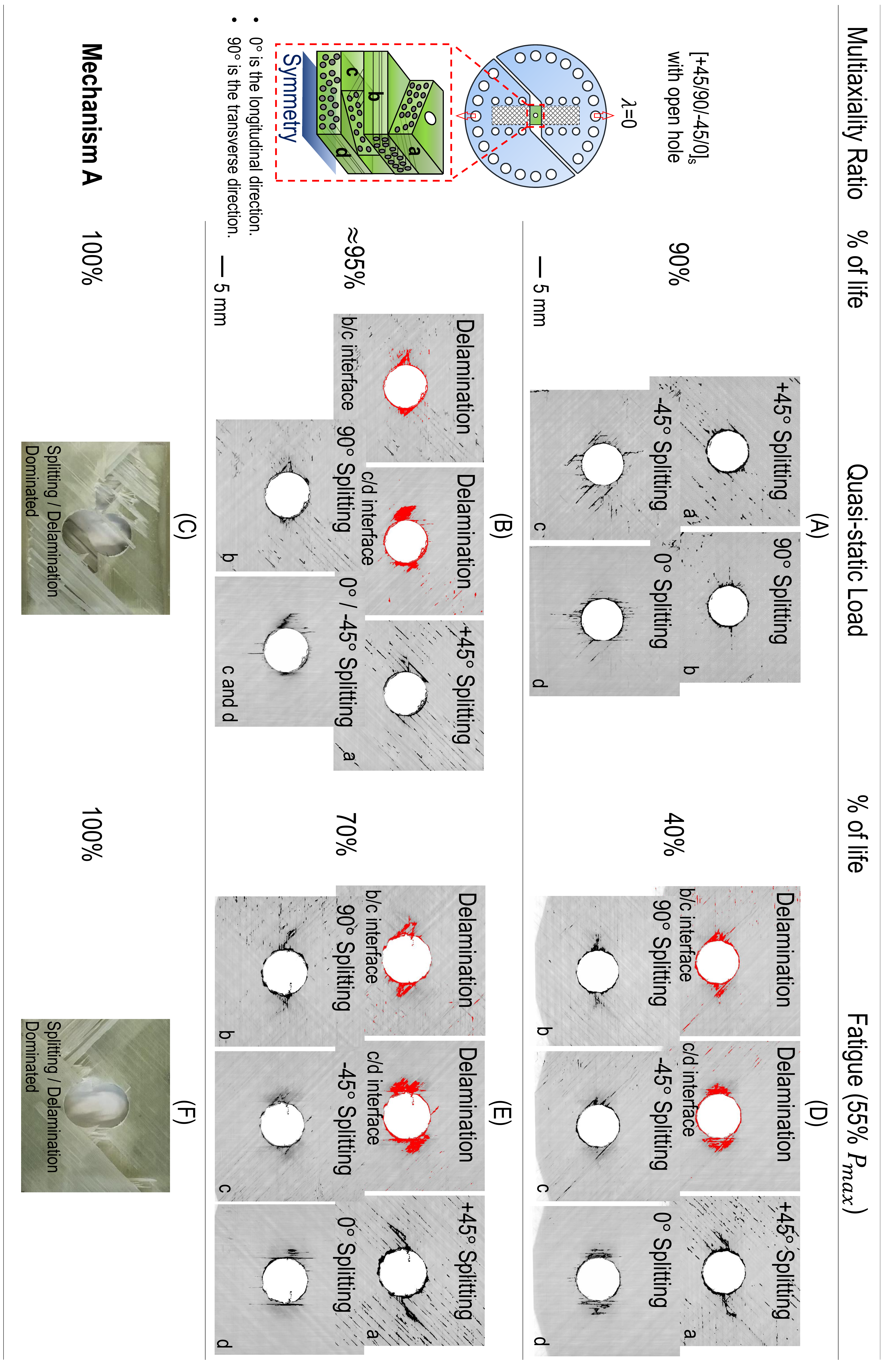}
\caption{Analysis of quasi-static and fatigue damage mechanisms by micro-computed tomography. Comparison between quasi-static and fatigue damage in $[+45/90/-45/0]_{s}$ specimens weakened by a 10 mm open hole for a multiaxiality ratio $\lambda=0$ in the $+45^{\circ}$,  $90^{\circ}$, $-45^{\circ}$, $0^{\circ}$ layers and at the interfaces. Note that the original colors of the images were inverted for a better visualization on the damage.}
\label{fig:damagemechanismquasiishole0}
\end{figure}

\leftline{(a) Mechanism A}
In the case of specimens subjected to tension-dominated loading conditions ($\lambda=0, 0.262$), the dominant mechanisms before the dramatic failure are the delamination between all the plies except for the middle plies and the significant splitting in $\pm$ $45^{\circ}$ plies. These damage is not the only contribution to the catastrophic failure and the breakage of $0^{\circ}$ piles also plays a pivotal role. As illustrated in Figure \ref{fig:damagemechanismquasiishole0}, the damage evolution as a function of fatigue lifetime for mechanism A can be summarized in the following:\vspace{0.4cm}

(1) splitting initiates in $90^{\circ}$ plies at notch tip (Fig. \ref{fig:damagemechanismquasiishole0}D); 

(2) splitting initiates in $0^{\circ}$ and $\pm$ $45^{\circ}$  plies at notch tip (Fig. \ref{fig:damagemechanismquasiishole0}D); 

(3) delamination between the foregoing plies initiates (Fig. \ref{fig:damagemechanismquasiishole0}D);

(4) splitting and delamination develop in/between the corresponding plies (Fig. \ref{fig:damagemechanismquasiishole0}E);

(5) splitting in $\pm$ $45^{\circ}$ plies grows significantly and delamination reaches critical 

condition accompanying the emergence of $0^{\circ}$ plies breakage (Fig. \ref{fig:damagemechanismquasiishole0}E-F).\vspace{0.4cm}

\begin{figure} [H]
\center
\includegraphics[scale=0.135]{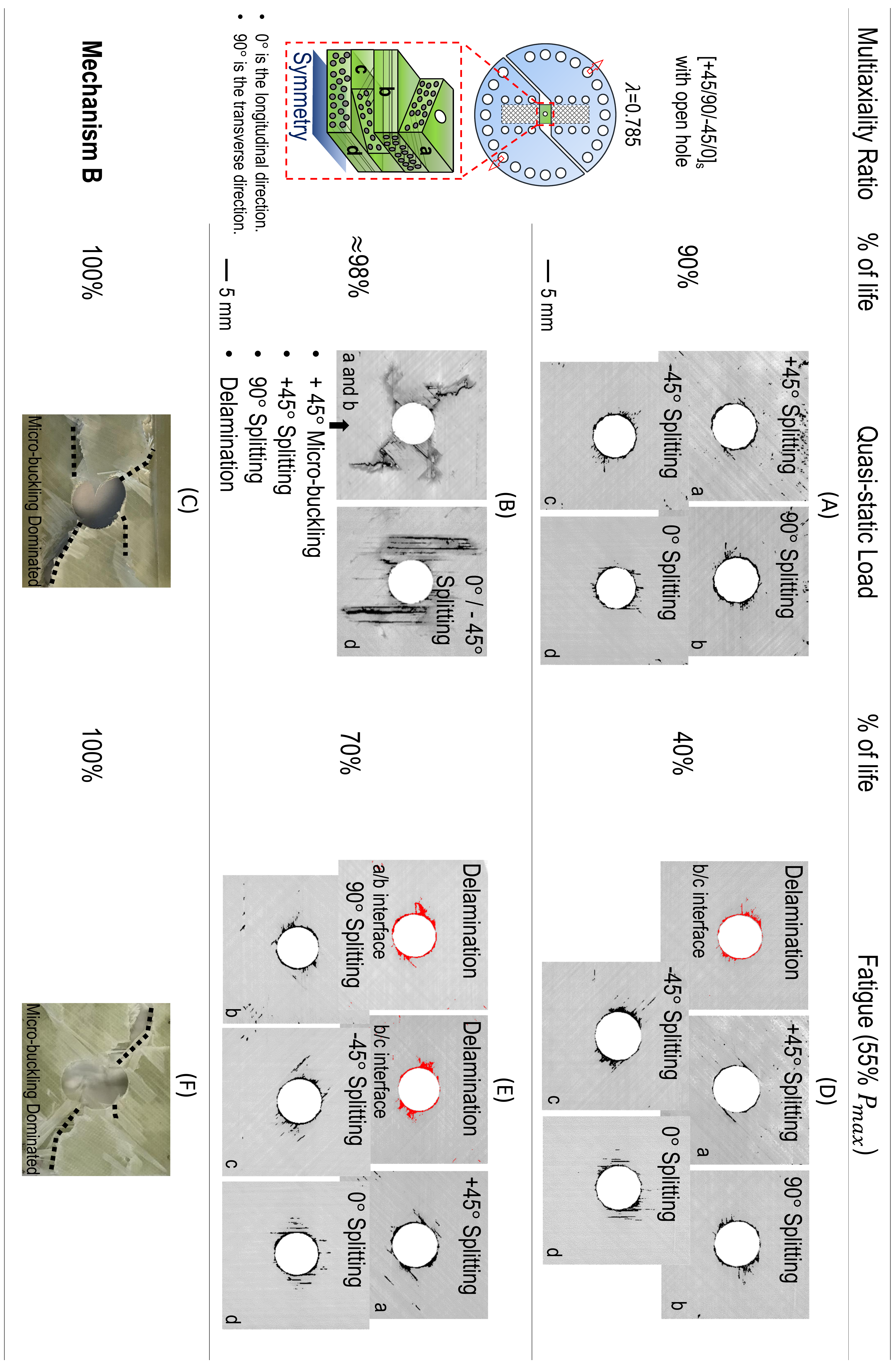}
\caption{Analysis of quasi-static and fatigue damage mechanisms by micro-computed tomography. Comparison between quasi-static and fatigue damage in $[+45/90/-45/0]_{s}$ specimens weakened by a 10 mm open hole for a multiaxiality ratio $\lambda=0.785$ in the $+45^{\circ}$,  $90^{\circ}$, $-45^{\circ}$, $0^{\circ}$ layers and at the interfaces. Note that the original colors of the images were inverted for a better visualization on the damage.}
\label{fig:damagemechanismquasiishole45}
\end{figure}

\leftline{(b) Mechanism B}
The specimens fail following Mechanism B when the shear load component is involved ($\lambda=0.785, 1.309$ and 1.571). In contrast to mechanism A, a significant reduced delamination between all the plies was observed. On the other hand, the direction of shear loading was established so that the  +$45^{\circ}$ plies are in compression while the -$45^{\circ}$ plies are in tension. Considering the fact that the compressive strength of an unidirectional ply is generally lower than its tensile strength, the dominant mechanism is the micro-buckling in +$45^{\circ}$ plies and the final failure is triggered by the significant growth of the micro-buckling due to the compression at the very late stage of the fatigue life. The paths of the micro-buckling usually start in a straight way at the two ends of the hole but not on the same line and then additional micro-buckling paths almost perpendicular to the +$45^{\circ}$ plies happen. These micro-buckling paths are usually in collaboration with the significant splitting in +$45^{\circ}$ plies as mentioned in the foregoing discussion of DIC analysis in Section \ref{sec:DIC}. Similar conclusions were drawn by Tan \emph{et al.} \cite{Tan,Tan2} on the quasi-static fracturing behavior of notched Carbon Fiber Reinforced Polymer (CFRP) laminates. As illustrated in Figures \ref{fig:damagemechanismquasiishole45}-\ref{fig:damagemechanismquasiishole90}, the damage evolution of mechanism B can be summarized in the following:\vspace{0.4cm}

(1) splitting initiates in $90^{\circ}$ plies at notch tip (Fig. \ref{fig:damagemechanismquasiishole45}D) ; 

(2) splitting initiates in $0^{\circ}$ and $\pm$ $45^{\circ}$ plies at notch tip (Fig. \ref{fig:damagemechanismquasiishole45}D); 

(3) a small amount of delamination between $90^{\circ}$ and $\pm$ $45^{\circ}$ plies starts to initiate (Fig. \ref{fig:damagemechanismquasiishole45}D);

(4) splitting and delamination develop in/between the corresponding plies (Fig. \ref{fig:damagemechanismquasiishole45}E);

(4) micro-buckling initiates in +$45^{\circ}$ plies at notch tip (Fig. \ref{fig:damagemechanismquasiishole45}F);

(5) micro-buckling in +$45^{\circ}$ plies grows unstably in collaboration with the significant 

splitting in +$45^{\circ}$ plies (Fig. \ref{fig:damagemechanismquasiishole45}F).

\subsubsection{Quantitative analysis: fatigue vs. quasi-static condition}
The similar quantitative analysis on the sub-critical damage as discussed in section \ref{sec:quantitativecrossply} was proceeded. As can be noted from Figure \ref{fig:damageevolution}g for the fatigue case, the specimen following mechanism A has larger delamination area compared to the one following mechanism B at 70\% of total fatigue life, with the highest value (1.13\%) for $\lambda=0$ and the lowest value (0.33\%) for $\lambda=1.571$. This is supported by the foregoing damage mechanisms showing that delamination takes less important role but micro-buckling dominates the fracturing behavior of notched quasi-isotropic laminates under shear-dominated loading. On the other hand, the evolution of total crack volume before 70\% of total fatigue life is close for each multiaxiality ratio. This explains the observation, reported in the previous work by Qiao \emph{et al.} \cite{Yaomulti}, that the stiffness deteriorates roughly 19 \% to 25 \% before catastrophic failure for various multiaxiality ratios. 

\begin{figure} [H]
\center
\includegraphics[scale=0.135]{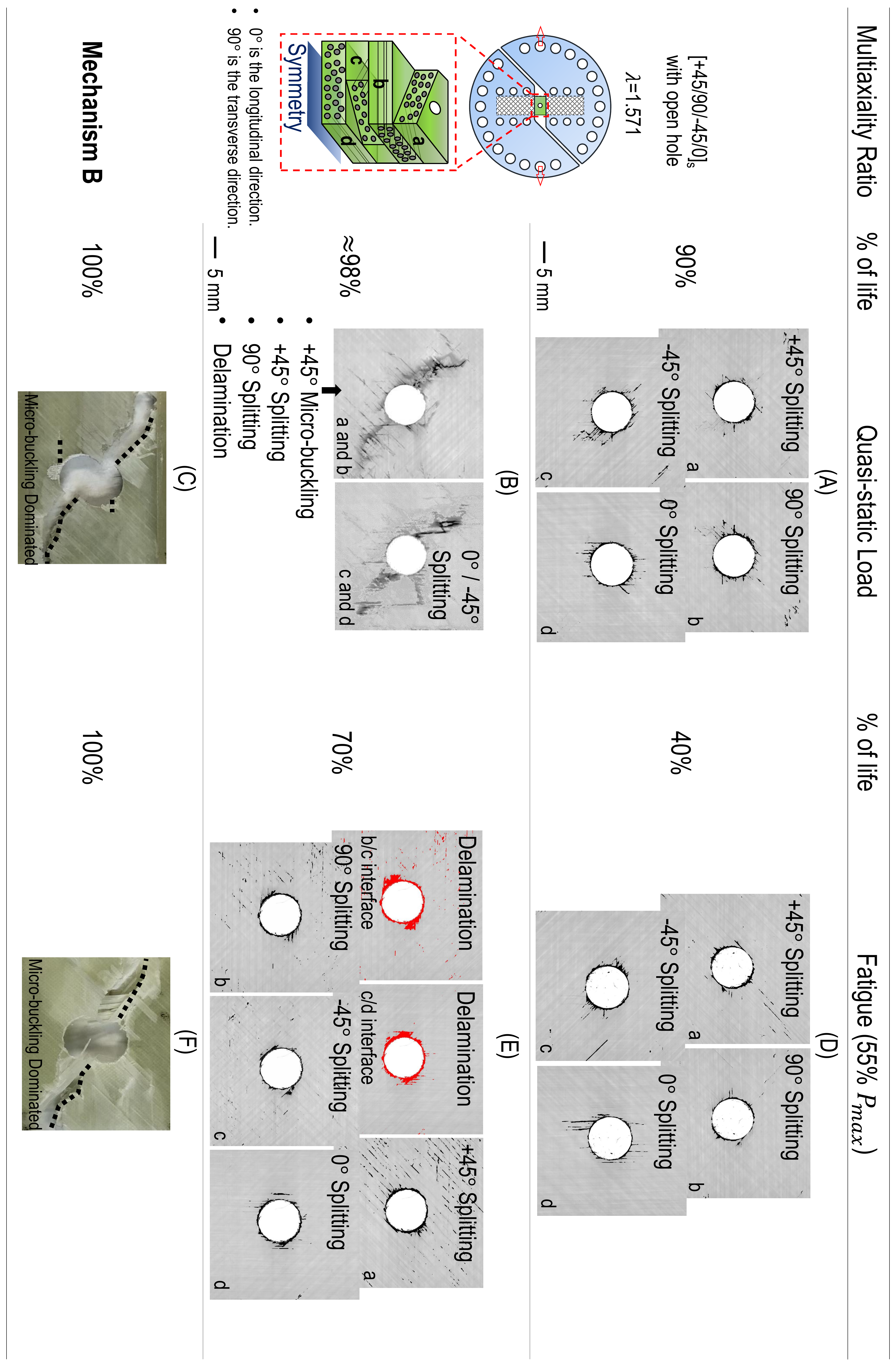}
\caption{Analysis of quasi-static and fatigue damage mechanisms by micro-computed tomography. Comparison between quasi-static and fatigue damage in $[+45/90/-45/0]_{s}$ specimens weakened by a 10 mm open hole for a multiaxiality ratio $\lambda=1.571$ in the $+45^{\circ}$,  $90^{\circ}$, $-45^{\circ}$, $0^{\circ}$ layers and at the interfaces. Note that the original colors of the images were inverted for a better visualization on the damage.}
\label{fig:damagemechanismquasiishole90}
\end{figure}

In the case of quasi-static loading, specimens follow the similar damage mechanisms for the foregoing fatigue case. However, in mechanism B, the straight paths of the micro-buckling in +$45^{\circ}$ plies at the two ends of the hole have a pronounced propagation with the increasing multiaxiality ratio as can be noted in Figures \ref{fig:damagemechanismquasiishole45}B and \ref{fig:damagemechanismquasiishole90}B . This difference is much more obvious in the quasi-static case compared to the fatigue case. Additionally, the overall sub-critical damage during quasi-static loading is also less diffused which is similar to the foregoing discussion on notched cross-ply laminates. As shown in Figures \ref{fig:damageevolution}e-h, total crack volume only has 2.7\% in average for different multiaxiality ratios and the highest delamination area has 0.13\% for multiaxiality ratio $\lambda=0.785$ at 90\% of total quasi-static life but these sub-critical damage grows rapidly close to the catastrophic failure.

\subsection{$[+45/90/-45/0]_{s}$ laminate with a central crack}
\begin{figure} [H]
\center
\includegraphics[scale=0.135]{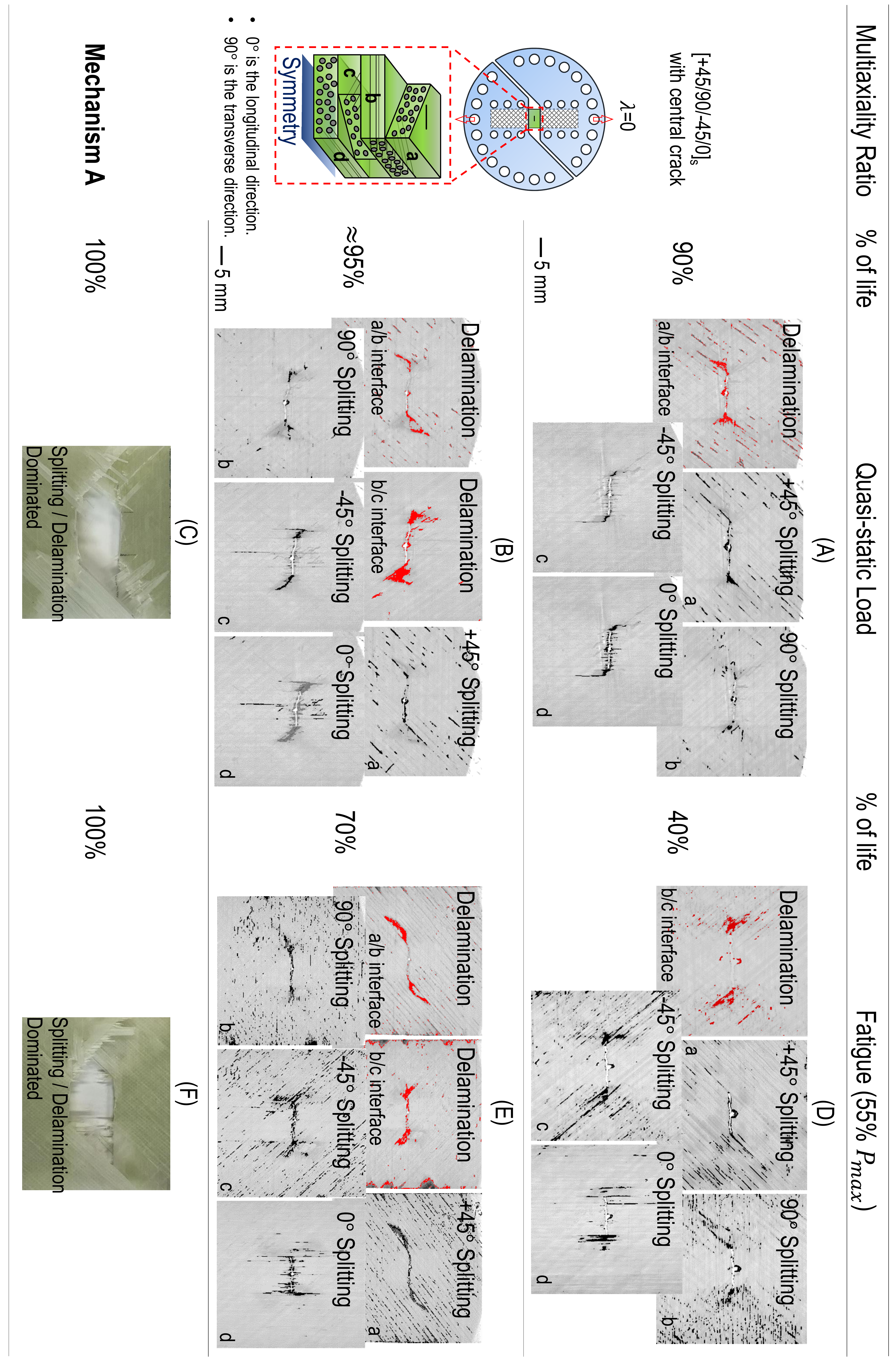}
\caption{Analysis of quasi-static and fatigue damage mechanisms by micro-computed tomography. Comparison between quasi-static and fatigue damage in $[+45/90/-45/0]_{s}$ specimens weakened by a 10 mm central crack for a multiaxiality ratio $\lambda=0$ in the $+45^{\circ}$,  $90^{\circ}$, $-45^{\circ}$, $0^{\circ}$ layers and at the interfaces. Note that the original colors of the images were inverted for a better visualization on the damage.}
\label{fig:damagemechanismquasiisocrack0}
\end{figure}

\begin{figure} [H]
\center
\includegraphics[scale=0.135]{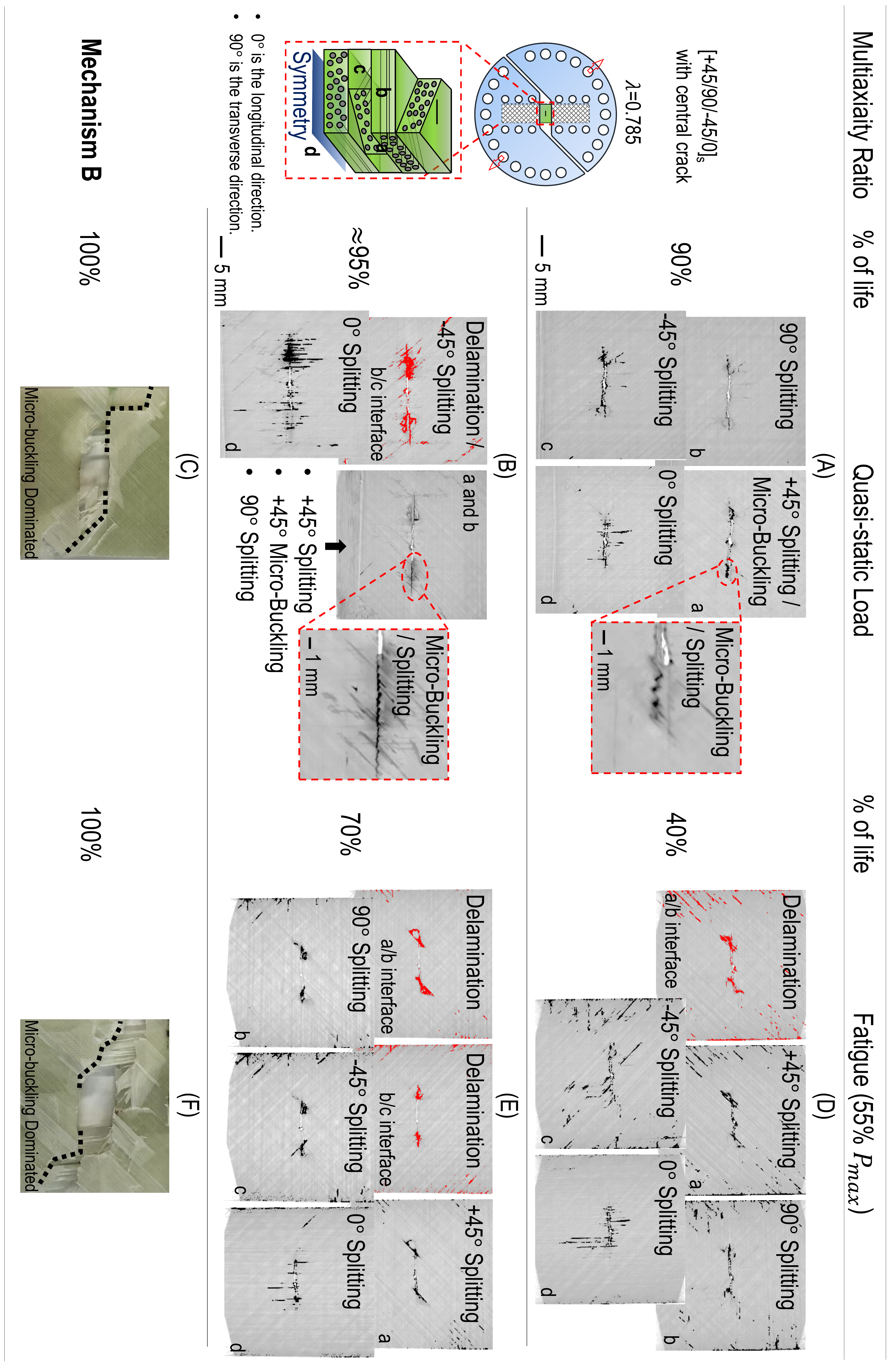}
\caption{Analysis of quasi-static and fatigue damage mechanisms by micro-computed tomography. Comparison between quasi-static and fatigue damage in $[+45/90/-45/0]_{s}$ specimens weakened by a 10 mm central crack for a multiaxiality ratio $\lambda=0.785$ in the $+45^{\circ}$,  $90^{\circ}$, $-45^{\circ}$, $0^{\circ}$ layers and at the interfaces. Note that the original colors of the images were inverted for a better visualization on the damage.}
\label{fig:damagemechanismquasiisocrack45}
\end{figure}

\begin{figure} [H]
\center
\includegraphics[scale=0.145]{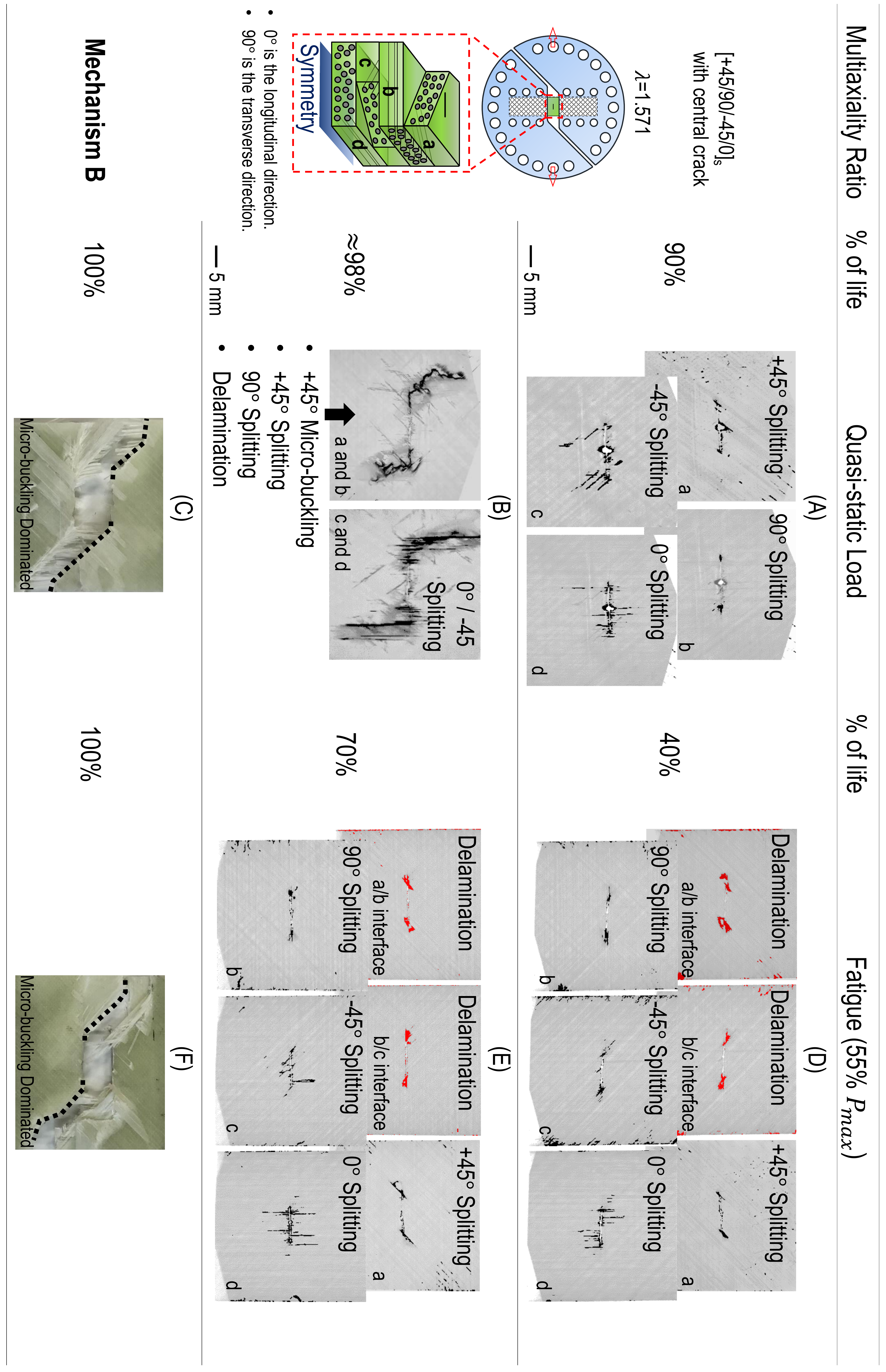}
\caption{Analysis of quasi-static and fatigue damage mechanisms by micro-computed tomography. Comparison between quasi-static and fatigue damage in $[+45/90/-45/0]_{s}$ specimens weakened by a 10 mm central crack for a multiaxiality ratio $\lambda=1.571$ in the $+45^{\circ}$,  $90^{\circ}$, $-45^{\circ}$, $0^{\circ}$ layers and at the interfaces. Note that the original colors of the images were inverted for a better visualization on the damage.}
\label{fig:damagemechanismquasiisocrack90}
\end{figure}

Having discussed the fracturing features and damage mechanisms for the quasi-isotropic specimens in presence of an open hole, similar damage mechanisms were observed for the same layup but featuring a central crack as shown in Figures \ref{fig:damagemechanismquasiisocrack0}-\ref{fig:damagemechanismquasiisocrack90}. The only difference in mechanism B (micro-buckling dominant) is the location of the micro-buckling paths ahead of the notch. As can be noted from Figures \ref{fig:damagemechanismquasiisocrack45}-\ref{fig:damagemechanismquasiisocrack90}, the straight micro-buckling paths in front of the notch are on the same line for multiaxiality ratios related to shear-dominated loading.

\subsubsection{Quantitative analysis: fatigue vs. quasi-isotropic condition}
In the fatigue case as shown in Figure \ref{fig:damageevolution}k, the larger delamination area was also observed for the specimen following mechanism A compared to the one following mechanism B before catastrophic fatigue failure. At 70\% of total fatigue life, the delamination reaches approximately 1.4\% for $\lambda=0$ whereas a significant reduction characterizes the other multiaxiality ratios, with 0.8\% for $\lambda=0.785$ and 0.55\% for $\lambda=1.571$. This is similar to the quasi-isotropic specimens featuring an open hole as discussed in the foregoing section. Another similarity was found in terms of the evolution of total crack volume confirming a similar gradual stiffness degradation throughout the fatigue life for various multiaxiality ratios. Despite these similarities, the specimens weakened by a central crack have a significant larger amount of total crack volume compared to the ones weakened by an open hole.

However, this does not occur for the quasi-static scenario which shows the central crack case having less amount of total crack volume. This indicates a different evolution of the Fracture Process Zone (FPZ) in quasi-static regime compared to fatigue which is significantly important for structural design since the previous study shows that specimens featuring a central hole behave better than the ones with the same layup but featuring a central crack in fatigue case but not for the quasi-static case \cite{Yaomulti}. It is worth mentioning again that less diffused sub-critical damage was also observed on the specimens featuring a central hole under quasi-static loading. As shown in Figure \ref{fig:damageevolution}i-l, total crack volume has roughly 1.96\% and the delamination area is significantly low (0.2\%) for different multiaxiality ratios at 90\% of total quasi-static life but these sub-critical damage also grows rapidly close to the catastrophic failure similar to the foregoing discussion on the quasi-static fracturing behavior.

\section{Conclusions}
Leveraging a synergistic combination of DIC and micro-computed tomography, this study investigated the fracturing morphology and the damage mechanisms of notched quasi-isotropic and cross-ply laminates under multi-axial loading. Based on the results obtained in this study, the following conclusions can be elaborated:

1. For the composite layups and specimen configurations investigated in this work, the sub-critical damage mechanisms under multi-axial fatigue were similar to the ones identified under quasi-static loading. However, it is worth mentioning that in notched quasi-isotropic laminates under shear-dominated loading a more pronounced straight micro-buckling paths ahead of the notch or hole was reported in the quasi-static case compared to the fatigue loading;

2. leveraging the $\mu$-CT results, it is shown that the damage mechanisms of the notched cross-ply laminates are dominated by splitting in $0^{\circ}$ plies under tension-dominated loading whereas a mix of delamination and fiber kinking in $0^{\circ}$ plies occurs with increasing shear load;

3. in the case of notched quasi-isotropic laminates, significant delamination and the splitting in $\pm$ $45^{\circ}$ plies characterizes the tension-dominated fatigue behavior whereas the main damage mechanism for the shear-dominated case is the micro-buckling in $+45^{\circ}$ plies in combination with splitting in the same plies;

4. The quantitative analysis of the crack volume via $\mu$-CT and of the maximum principal strain distribution by means of DIC reveals a substantial difference in the damage evolution between the quasi-static and fatigue loading conditions. In fatigue, the sub-critical damage is distributed across a larger volume compared to the quasi-static case, leading to a significant strain redistribution. Further, the rate of crack volume increase throughout the life of the specimen is substantially different. While in the quasi-static case the crack volume increases quickly at almost a constant rate, in fatigue it exhibits a slow change followed by an abrupt increase towards final failure;

5. the foregoing results are of utmost importance for the structural design of polymer matrix composites under multi-axial loading condition but so far rarely investigated. This study provides a comprehensive damage evaluation on the fracturing behavior of notched laminates which can be used to validate the existing models for the design of composite structures under multi-axial stress states.    



 % \bibliographystyle{elsarticle-num}
 % \bibliography{carbonbib}

\section*{Acknowledgments}
Marco Salviato acknowledges the financial support from the Haythornthwaite Foundation through the ASME Haythornthwaite Young Investigator Award and from the University of Washington Royalty Research Fund. The work was also partially supported by NSF CMMI 1428436 ``MRI: Acquisition of a 3D X-Ray Computed Tomography Scanner for Imaging of Large Size Infrastructure, Biological, and Mechanical Components" awarded to the University of Washington.

\clearpage

\section*{Figures and Tables}

\newpage

\newpage

\newpage

\newpage

\newpage

\newpage

\end{document}